\documentclass[showpacs,preprintnumbers,amsmath,amssymb,nofootinbib,floatfix]{revtex4}

\usepackage[dvips]{graphicx}
\usepackage{dcolumn}
\usepackage{bm}

\def\mapgeq{\mathbin{\lower.3ex\hbox{$\buildrel>\over{\smash{\scriptstyle\sim}\vphantom{_x}}$}}}
\def\mapleq{\mathbin{\lower.3ex\hbox{$\buildrel<\over{\smash{\scriptstyle\sim}\vphantom{_x}}$}}}
\def\mapgeqeq{\mathbi{\lower.3ex\hbox{$\buildrel>\over{\smash{\scriptstyle\approx}\vphantom{_2}}$}}}
\def\mapleqeq{\mathbin{\lower.3ex\hbox{$\buildrel<\over{\smash{\scriptstyle\approx}\vphantom{_2}}$}}}
\mathchardef\hanaO="724F

\def\Journal#1#2#3#4{{#1} {\bf #2} (#4) #3}
\def\MPL{Mod. Phys. Lett. A}

\def\NPB{Nucl. Phys. B}

\def\NPSUPPL{Nucl. Phys. Proc. Suppl.}
\def\PLB{{Phys. Lett.} B}

\def\PRL{Phys. Rev. Lett.}
\def\RMP{Rev. Mod. Phys.}
\def\PRD{Phys. Rev. D}

\def\PTP{Prog. Theor. Phys.}
\def\JHEP{JHEP}

\def\NPBSUPPL{Nucl. Phys. B. Proc. Suppl.}
\def\EPJ{Euro. Phys. J. C}

\def\JETPUSSR{Sov. Phys. JETP}

\def\ZETP{Zh. Eksp. Teor. Fiz.}

\def\JPG{J. Phys. G}

\def\APJ{Astrophys. J.}

\def\NJP{New J. Phys.}

\def\Erratum{Erratum-ibid}
\def\SCIENCE{Science}

\begin{document}


 \title{Generalized Scaling in Flavor Neutrino Masses}

\author{Masaki Yasu\`{e}}
\email{yasue@keyaki.cc.u-tokai.ac.jp}
\affiliation{\vspace{3mm}%
\sl Department of Physics, Tokai University,\\
4-1-1 Kitakaname, Hiratsuka, Kanagawa 259-1292, Japan\\
}

\date{September, 2012}

\begin{abstract}
Scaling in flavor neutrino masses $M_{ij}$ ($i,j$=$e,\mu,\tau$) can be described by two angles: $\theta_{SC}$ and the atmospheric neutrino mixing angle $\theta_{23}$.  For $A$=${\cos ^2}{\theta _{SC}}+{\sin ^2}{\theta _{SC}}t_{23}^4$ and $B$=${\cos ^2}{\theta _{SC}}-{\sin ^2}{\theta _{SC}}t_{23}^2$, where $t_{23}=\tan\theta_{23}$, our scaling ansatz dictates that $M_{i\tau }/M_{i\mu }$ = $- \kappa _it_{23}$ ($i$=$e,\mu,\tau$) with $\kappa _e$=1, $\kappa _\mu$=$B/A$ and $\kappa _\tau$=$1/B$ and leads to the vanishing reactor neutrino mixing angle $\theta_{13}=0$.  This generalized scaling is naturally realized in seesaw textures. To obtain $\theta_{13}\neq 0$ as required by the recent experimental results, we introduce breaking terms of scaling ansatz, which are taken to keep $M_{\mu\tau }/M_{\mu\mu }$ = $- \kappa _\mu t_{23}$ intact even at $\theta_{13}\neq 0$.  We derive relations that connect CP violating phases with phases of flavor neutrino masses, which are found to be numerically supported. The angle $\theta_{SC}$ is observed to be $0.91 \lesssim\sin^2\theta_{SC}\lesssim 0.93$ for the normal mass hierarchy and $\sin^2\theta_{SC}\lesssim 0.33$ for the inverted mass hierarchy.  Also observed is the size of $\left|M_{ee}\right|$ to be measured in neutrinoless double beta decay, which is 0.001-0.004 eV (0.02 eV-0.05 eV) in the normal (inverted) mass hierarchy. 
\end{abstract}

\maketitle
\section{\label{sec:intro} Introduction}
Three flavor neutrinos $\nu_{e,\mu,\tau}$ are mixed into three massive neutrinos $\nu_{1,2,3}$ during their flight.  The result of these mixings have been observed as the phenomena of neutrino oscillations \cite{PMNS}. Various experiments have detected the $\nu_\mu$-$\nu_\tau$ mixing via the atmospheric and accelerator neutrino oscillations \cite{atmospheric,accelerator}, the $\nu_e$-$\nu_\mu$ mixing via the solar and reactor neutrino oscillations \cite{oldsolar,solar,reactor} and the $\nu_e$-$\nu_\tau$ mixing via the reactor neutrino oscillation \cite{theta13}.  All these oscillations have indicated that neutrinos have extremely small masses much smaller than the electron mass \cite{NuData,NuData2}. Since the nonvanishing $\nu_e$-$\nu_\tau$ mixing has been observed to occur, Dirac CP-violation is expected to cause sizable effects in neutrino mixings. Effects of CP-violation can be described by phases of the Pontecorvo-Maki-Nakagawa-Sakata (PMNS) unitary matrix $U_{PMNS}$ \cite{PMNS}, which converts $\nu_{1,2,3}$ into $\nu_{e,\mu,\tau}$.  Namely, $U_{PMNS}$ has three phases, one CP-violating Dirac phase $\delta$ and two CP-violating Majorana phases $\phi_{2,3}$ \cite{CPViolation} and is given by $U_{PMNS}=U^0_\nu K^0$ \cite{PDG} with
\begin{eqnarray}
U^0_\nu&=&\left( \begin{array}{ccc}
  c_{12}c_{13} &  s_{12}c_{13}&  s_{13}e^{-i\delta}\\
  -c_{23}s_{12}-s_{23}c_{12}s_{13}e^{i\delta}
                                 &  c_{23}c_{12}-s_{23}s_{12}s_{13}e^{i\delta}
                                 &  s_{23}c_{13}\\
  s_{23}s_{12}-c_{23}c_{12}s_{13}e^{i\delta}
                                 &  -s_{23}c_{12}-c_{23}s_{12}s_{13}e^{i\delta}
                                 & c_{23}c_{13}\\
\end{array} \right),
\nonumber \\
K^0 &=& {\rm diag}(1, e^{i\phi_2/2}, e^{i\phi_3/2}),
\label{Eq:UuPDG}
\end{eqnarray}
where $c_{ij}=\cos\theta_{ij}$ and $s_{ij}=\sin\theta_{ij}$ and $\theta_{ij}$ represents a $\nu_i$-$\nu_j$ mixing angle ($i,j$=1,2,3). The experimental observation of the parameters in $U^0_\nu$ for the normal mass hierarchy \cite{NuData} shows that 
\begin{eqnarray}
\Delta m^2_{21} ~[10^{-5}~{\rm eV}^2] & = &7.62\pm 0.19,
\quad
\Delta m^2_{31} ~[10^{-3}~{\rm eV}^2] = 2.55
{\footnotesize
{\begin{array}{*{20}c}
   { + 0.06}  \\
   { - 0.09}  \\
\end{array}}
},
\label{Eq:NuDataMass}\\
\sin ^2 \theta _{12} & = &0.320
{\footnotesize
{\begin{array}{*{20}c}
   { + 0.016}  \\
   { - 0.017}  \\
\end{array}}
},
\quad
\sin ^2 \theta _{23} =0.427
{\footnotesize
{\begin{array}{*{20}c}
   { + 0.034}  \\
   { - 0.027}  \\
\end{array}}
}
~\left(
0.613
{\footnotesize
{\begin{array}{*{20}c}
   { + 0.022}  \\
   { - 0.040}  \\
\end{array}}
}
\right),
\quad
\sin ^2 \theta _{13} =0.0246
{\footnotesize
 {\begin{array}{*{20}c}
   { + 0.0029}  \\
   { - 0.0028}  \\
\end{array}}
},
\label{Eq:NuDataAngle}\\
\frac{\delta _{PC}}{\pi} &=& 0.80
{\footnotesize
 {\begin{array}{*{20}c}
   { + 1.20}  \\
   { - 0.80}  \\
\end{array}}
},
\label{Eq:NuDataPhase}
\end{eqnarray}
where $\Delta m^2_{ij}=m^2_i-m^2_j$ and $m_i$ stands for a mass of $\nu_i$ ($i=1,2,3$).  For the inverted mass hierarchy ($\Delta m^2_{31}<0$), the results are not so different from Eqs.(\ref{Eq:NuDataMass})-(\ref{Eq:NuDataPhase}).  There is another similar analysis that has reported the slightly smaller values of $\sin^2\theta_{23} = 0.365-0.410$ \cite{NuData2}.

The experimental results show $\sin^2\theta_{13}\ll 1$ suggesting the ideal mixing of $\theta_{13} = 0$, which can be realized by a certain theoretical assumption on the neutrino mixing such as the bimaximal mixing \cite{Bimaximal}, the tribimaximal mixing \cite{Tribimaximal}, the transposed tribimaximal mixing \cite{transposedTribimaximal}, the bipair mixing \cite{Bipair}, the golden ratio scheme \cite{goldenRatio}, the hexagonal mixing \cite{hexagonal}, the $\mu$-$\tau$ symmetry \cite{mu-tau} and the scaling ansatz \cite{Scaling}. In this article, we would like to discuss another theoretical possibility to lead to $\theta_{13}=0$, which corresponds to the generalization of the scaling rule of Ref.\cite{Scaling}.  Our generalized scaling ansatz utilizes one angle to be denoted by $\theta_{SC}$, whose origin can be traced back to seesaw textures that predict $\theta_{13}=0$. It turns out that the original scaling ansatz is recovered at $\theta_{SC}=0$.  In the next section Sec.\ref{sec:scaling}, we define the generalized scaling rule and discuss its connection with seesaw textures.  In Sec.\ref{sec:CPviolation}, we introduce the breaking of the generalized scaling ansatz to generate the nonvanishing $\theta_{13}$ and develop theoretical arguments to describe CP-violations.  In Sec.\ref{sec:calculation}, we perform numerical analysis to evaluate effects of CP-violation and to find how CP-violation is controlled by phases of flavor neutrino masses. Our numerical results are compared with theoretical expectations. The final section Sec.\ref{sec:summary} is devoted to summary and discussions.

\section{\label{sec:scaling} Generalized Scaling}
To obtain $\theta_{13} = 0$, flavor neutrino masses to be denoted by $M_{ij}$ ($i,j$=$e,\mu,\tau$) should satisfy the following constraints \cite{theta_13=0}:
\begin{eqnarray}
{M_{e\tau }} &=& - {t_{23}}{M_{e\mu }},
\label{Eq:Constraints-Metau}\\
{M_{\tau \tau }} &=& {M_{\mu \mu }} + \frac{{1 - t_{23}^2}}{{{t_{23}}}}{M_{\mu \tau }}.
\label{Eq:Constraints-Mtautau}
\end{eqnarray}
The simplest solution to Eq.(\ref{Eq:Constraints-Mtautau}) can be obtained by using an angle $\theta_{SC}$ and turns out to be
\begin{eqnarray}
\frac{{{M_{\mu \mu }}}}{{{M_{\tau \tau }}}} &=&\frac{{{{\cos }^2}{\theta _{SC}}}}{{t_{23}^2}} + {\sin ^2}{\theta _{SC}}t_{23}^2,
\nonumber\\
\frac{{{M_{\mu \tau }}}}{{{M_{\tau \tau }}}}&=& - \left( {\frac{{{{\cos }^2}{\theta _{SC}}}}{{{t_{23}}}} - {{\sin }^2}{\theta _{SC}}{t_{23}}} \right).
\label{Eq:Scaling}
\end{eqnarray}
These relations provide the following scaling rule:
\begin{eqnarray}
\frac{{{M_{i\tau }}}}{{{M_{i\mu }}}} =  - {\kappa _i}{t_{23}}~\left(i=e,\mu,\tau\right),
\label{Eq:ScalingRule}
\end{eqnarray}
where $\left(\kappa_e, \kappa_\mu, \kappa_\tau\right)$=$\left( 1,B/A,1/B\right)$, and 
\begin{eqnarray}
A = {\cos ^2}{\theta _{SC}} + {\sin ^2}{\theta _{SC}}t_{23}^4,
\quad
B = {\cos ^2}{\theta _{SC}} - {\sin ^2}{\theta _{SC}}t_{23}^2.
\label{Eq:A-B}
\end{eqnarray}
Our generalized scaling ansatz is described by Eq.(\ref{Eq:ScalingRule}).  The flavor neutrino texture satisfying Eq.(\ref{Eq:ScalingRule}) is regarded as a new texture giving $\theta_{13}=0$.  The angle $\theta_{SC}$ is defined from ${{{M_{\mu\tau }}}}/{{{M_{\mu\mu }}}} =  - {\kappa _\mu}{t_{23}}$ to be:
\begin{eqnarray}
{{\sin }^2}{\theta _{SC}} = \frac{{c_{23}^2\left( {{M_{\mu \tau }} + {t_{23}}{M_{\mu \mu }}} \right)}}{{\left( {1 - t_{23}^2} \right){M_{\mu \tau }} + {t_{23}}{M_{\mu \mu }}}}.
\label{Eq:Sin2SC}
\end{eqnarray}

The origin of $\theta_{SC}$ can be found in a seesaw model \cite{SeeSaw}, which has $M_{1,2,3}$ as diagonal masses of three heavy neutrinos and a 3$\times$3 Dirac neutrino mass matrix $m_D$ defined by
\begin{eqnarray}
{m_D} = \left( {\begin{array}{*{20}{c}}
{{a_1}}&{{b_1}}&{{c_1}}\\
{{a_2}}&{{b_2}}&{{c_2}}\\
{{a_3}}&{{b_3}}&{{c_3}}
\end{array}} \right),
\label{Eq:Seesaw}
\end{eqnarray}
on the basis that masses of charged leptons are diagonal.  Since
\begin{eqnarray}
{M_{e\mu }} &=& -\left({\frac{{{a_1}{a_2}}}{{{M_1}}} + \frac{{{b_1}{b_2}}}{{{M_2}}} + \frac{{{c_1}{c_2}}}{{{M_3}}}}\right),
\quad
{M_{e\tau }} = -\left({\frac{{{a_1}{a_3}}}{{{M_1}}} + \frac{{{b_1}{b_3}}}{{{M_2}}} + \frac{{{c_1}{c_3}}}{{{M_3}}}}\right),
\label{Eq:Mei-Seesaw}\\
{M_{\mu \mu }} &=& -\left({\frac{{a_2^2}}{{{M_1}}} + \frac{{b_2^2}}{{{M_2}}} + \frac{{c_2^2}}{{{M_3}}}}\right),
\quad
{M_{\mu \tau }} = -\left({\frac{{{a_2}{a_3}}}{{{M_1}}} + \frac{{{b_2}{b_3}}}{{{M_2}}} + \frac{{{c_2}{c_3}}}{{{M_3}}}}\right),
\quad
{M_{\tau \tau }} = -\left({\frac{{a_3^2}}{{{M_1}}} + \frac{{b_3^2}}{{{M_2}}} + \frac{{c_3^2}}{{{M_3}}}}\right).
\label{Eq:Mij-Seesaw}
\end{eqnarray}
Since the generalized scaling rule is derived as a solution to Eqs.(\ref{Eq:Constraints-Metau}) and (\ref{Eq:Constraints-Mtautau}), its seesaw version can also be found as their solution. The conditions Eqs.(\ref{Eq:Constraints-Metau}) and (\ref{Eq:Constraints-Mtautau}) are, respectively, converted into 
\begin{eqnarray}
&&
{\frac{{{a_1}\left( {{a_3} + {t_{23}}{a_2}} \right)}}{{{M_1}}} + \frac{{{b_1}\left( {{b_3} + {t_{23}}{b_2}} \right)}}{{{M_2}}} + \frac{{{c_1}\left( {{c_3} + {t_{23}}{c_2}} \right)}}{{{M_3}}} = 0},
\label{Eq:Constraints-Metau-Seesaw}\\
&&
{\frac{{\left( {{t_{23}}{a_3} - {a_2}} \right)\left( {{a_3} + {t_{23}}{a_2}} \right)}}{{{M_1}}} + \frac{{\left( {{t_{23}}{b_3} - {b_2}} \right)\left( {{b_3} + {t_{23}}{b_2}} \right)}}{{{M_2}}} + \frac{{\left( {{t_{23}}{c_3} - {c_2}} \right)\left( {{c_3} + {t_{23}}{c_2}} \right)}}{{{M_3}}} = 0}.
\label{Eq:Constraints-Mtautau-Seesaw}
\end{eqnarray}
To see the definition of $\theta_{SC}$ in terms of seesaw textures, it is sufficient to use $a_1=0$ as one of the solutions to Eqs.(\ref{Eq:Constraints-Metau-Seesaw}) and (\ref{Eq:Constraints-Mtautau-Seesaw}).  The generalized scaling anzatz is recovered by the following two types of solutions consisting of (a) $a_3 = -t_{23}a_2$ or (b) $a_3 = a_2/t_{23}$ provided that $a_1={b_3} + {t_{23}}{b_2} = {c_3} + {t_{23}}{c_2} = 0$.  The type (a) solution corresponds to
\begin{eqnarray}
&&
{m_D} = \left( {\begin{array}{*{20}{c}}
0&{{b_1}}&{{c_1}}\\
{{a_2}}&{{b_2}}&{{c_2}}\\
{ - {t_{23}}{a_2}}&{ - {t_{23}}{b_2}}&{ - {t_{23}}{c_2}}
\end{array}} \right),
\label{Eq:Solution-MSeesaw1}
\end{eqnarray}
from which Eq.(\ref{Eq:Sin2SC}) yields
\begin{eqnarray}
{\sin ^2}{\theta _{SC}} = 0.
\label{Eq:Solution-SC-Mtautau-Seesaw1}
\end{eqnarray}
On the other hand, the type (b) solution corresponds to
\begin{eqnarray}
&&
{m_D} = \left( {\begin{array}{*{20}{c}}
0&{{b_1}}&{{c_1}}\\
{{a_2}}&{{b_2}}&{{c_2}}\\
{{{a_2}}/{{t_{23}}}}&{ - {t_{23}}{b_2}}&{ - {t_{23}}{c_2}}
\end{array}} \right),
\label{Eq:Solution-MSeesaw2}
\end{eqnarray}
from which Eq.(\ref{Eq:Sin2SC}) yields
\begin{eqnarray}
\sin ^2\theta _{SC} = - \frac{1}{{{M_{\tau \tau }}}}\frac{{a_3^2}}{{{M_1}}},
\label{Eq:Solution-SC-Mtautau-Seesaw2}
\end{eqnarray}
where
\begin{eqnarray}
\arg \left( {\frac{{a_3^2}}{{{M_1}}}} \right) = \arg \left( {\frac{{b_3^2}}{{{M_2}}} + \frac{{c_3^2}}{{{M_3}}}} \right)~(\rm mod~\pi),
\label{Eq:Solution-SC-Arg-Seesaw2}
\end{eqnarray}
should be fulfilled.  There are other similar solutions such as those involving $b_1=0$ or $c_1=0$ instead of $a_1=0$. Therefore, the origin of $\theta_{SC}$ is linked to the existence of two types of solutions to Eqs.(\ref{Eq:Constraints-Metau-Seesaw}) and (\ref{Eq:Constraints-Mtautau-Seesaw}), which are either $a_3 = a_2/t_{23}$ or $x_3 = -t_{23} x_2$ ($x=b,c$), ensuring $\theta_{13}=0$, and tells us that $\sin ^2\theta _{SC}\propto a^2_3$ in the present example.

The constraint on $\theta_{SC}$ arises from $m^2_2 > m^2_1$.  The neutrino masses are calculated from  
\begin{eqnarray}
&&
{\tilde m}_1 \left( \equiv m_1e^{-i\varphi_1}\right)= \frac{1}{{c_{12}^2 - s_{12}^2}}\left( {c_{12}^2{M_{ee}} - s_{12}^2\frac{m}{{c_{23}^2}}} \right),
\nonumber\\
&&
{\tilde m}_2 \left( \equiv m_2e^{-i\varphi_2}\right)=  - \frac{1}{{c_{12}^2 - s_{12}^2}}\left( {s_{12}^2{M_{ee}} - c_{12}^2\frac{m}{{c_{23}^2}}} \right),
\nonumber\\
&&
{\tilde m}_3\left( \equiv m_3e^{-i\varphi_3}\right) = \frac{{{M_{\mu \mu }} - m}}{{s_{23}^2}},
\label{Eq:MassesIdeal}\\
&&
\tan {\theta _{23}} =  - \frac{{{M_{e\tau }}}}{{{M_{e\mu }}}},
\quad
\tan 2{\theta _{12}} = \frac{{2{c_{23}}{M_{e\mu }}}}{{m - c_{23}^2{M_{ee}}}},
\label{Eq:AnglesIdeal}
\end{eqnarray}
where $\varphi _i$ ($i=1,2,3$) are three Majorana phases giving $\phi_{2,3}=\varphi_{2,3}-\varphi_1$ and
\begin{eqnarray}
m = \frac{{{{\cos }^2}{\theta _{SC}}}}{{{{\cos }^2}{\theta _{SC}} + {{\sin }^2}{\theta _{SC}}t_{23}^4}}{M_{\mu \mu }}.
\label{Eq:MassRef}
\end{eqnarray}
It is obvious that $m_3=0$ for $\sin^2\theta_{SC}=0$.  The original scaling ansatz \cite{Scaling} is, thus, recovered by $\sin^2\theta_{SC}=0$.  The constraint of $m^2_2>m^2_1$ requires that
\begin{eqnarray}
{\left| {\frac{{{{\cos }^2}{\theta _{SC}}}}{{{{\cos }^2}{\theta _{SC}} + {{\sin }^2}{\theta _{SC}}t_{23}^4}}{M_{\mu \mu }}} \right| > \left| {c_{23}^2{M_{ee}}} \right|},
\label{Eq:m2-m1}
\end{eqnarray}
which excludes the case with $\cos^2\theta_{SC}=0$.

\section{\label{sec:CPviolation} Minimal Breaking and CP-Violation}
To discuss Dirac CP-violation, we have to obtain $\theta_{13} \neq 0$ and include breaking terms of the general scaling anzatz. Considering 
Eqs.(\ref{Eq:Constraints-Metau}) and (\ref{Eq:Constraints-Mtautau}), we introduce the following breaking terms supplied by
\begin{eqnarray}
\delta M_{e \tau } &=& M_{e\tau } + t_{23}M_{e\mu },
\nonumber\\
\delta M_{\tau \tau } &=& {{M_{\tau \tau }} - \left( {{M_{\mu \mu }} + \frac{{1 - t_{23}^2}}{{{t_{23}}}}{M_{\mu \tau }}} \right)},
\label{Eq:deltaM,}
\end{eqnarray}
which parameterize a neutrino mass matrix $M_\nu$ as follows:
\begin{eqnarray}
{M_\nu } = \left( {\begin{array}{*{20}{c}}
{{M_{ee}}}&{{M_{e\mu }}}&{{M_{e\tau }}}\\
{{M_{e\mu }}}&{{M_{\mu \mu }}}&{{M_{\mu \tau }}}\\
{{M_{e\tau }}}&{{M_{\mu \tau }}}&{{M_{\tau \tau }}}
\end{array}} \right) = M_{scaling} + \delta M
\label{Eq:MassMatrix}
\end{eqnarray}
with
\begin{eqnarray}
M_{scaling} &=& \left( {\begin{array}{*{20}{c}}
{{M_{ee}}}&{{M_{e\mu }}}&{ - {t_{23}}{M_{e\mu }}}\\
{{M_{e\mu }}}&{{M_{\mu \mu }}}&{ - \frac{B}{A}{t_{23}}{M_{\mu \mu }}}\\
{ - {t_{23}}{M_{e\mu }}}&{ - \frac{B}{A}{t_{23}}{M_{\mu \mu }}}&{\frac{1}{A}t_{23}^2{M_{\mu \mu }}}
\end{array}} \right),
\nonumber\\
\delta M &=& \left( {\begin{array}{*{20}{c}}
0&0&\delta M_{e\tau }\\
0&0&0\\
\delta M_{e\tau }&0&\delta M_{\tau \tau }
\end{array}} \right).
\label{Eq:MassMatrixBreaking}
\end{eqnarray}
These breaking terms always generate the nonvanishing $\theta_{13}$ and necessarily induce Dirac CP-violation.  It should be noted that there are other breakings that satisfy Eq.(\ref{Eq:Constraints-Mtautau}).  The nonvanishing $\theta_{13}$ cannot be induced by this type of breakings, which can be parameterized by
\begin{eqnarray}
\Delta M &=& \left( {\begin{array}{*{20}{c}}
0&0&0\\
0&0&{\Delta M_{\mu \tau }}\\
0&{\Delta M_{\mu \tau }}&{\Delta M_{\tau \tau }}
\end{array}} \right),
\label{Eq:DeltaM_theta13=0}
\end{eqnarray}
where
\begin{eqnarray}
\Delta M_{\mu \tau } &=& {M_{\mu \tau }} + \frac{B}{A}{t_{23}}{M_{\mu \mu }},
\nonumber\\
\Delta M_{\tau \tau } &=& \left( {1 - \frac{{t_{23}^2}}{A}} \right){M_{\mu \mu }} + \frac{{1 - t_{23}^2}}{{{t_{23}}}}{M_{\mu \tau }} = \frac{{1 - t_{23}^2}}{{{t_{23}}}}\Delta M_{\mu \tau }.
\label{Eq:DeltaM}
\end{eqnarray}
It turns out that the contribution to $\delta M_{\tau\tau}$ from $\Delta M_{\mu\tau,\tau\tau}$ vanishes. As long as the breaking terms are restricted to be $\delta M_{e\tau,\tau\tau}$, we are implicitly assuming that $\Delta M_{\mu\tau}=0$.  The breaking due to $\delta M_{e\tau,\tau\tau}$ only can be regarded as a minimal one because the other breaking due to $\Delta M_{\mu\tau,\tau\tau}$ is not included.  It is understood that
\begin{itemize}
\item the minimal breaking is the breaking that requires $\Delta M_{\mu\tau}=0$.
\end{itemize}
Because of this requirement, Eq.(\ref{Eq:Sin2SC}) still defines $\theta_{SC}$  even if $\theta_{13}\neq 0$.

To see what the implicit condition $\Delta M_{\mu\tau}=0$ suggests on the current argument, we use the diagonal mass matrix $M_{mass}$ = diag.$(m_1, m_2, m_3)$ applied to $M_\nu = U_{PMNS}^\ast M_{mass}U_{PMNS}^\dagger$ and find that one of the neutrino masses, which is taken to be $m_2$, is determined as follows:
\begin{eqnarray}
{{\tilde m}_2} = \frac{{\left[ {\left( {{c_{23}}{s_{12}} + {s_{23}}{c_{12}}\tilde s_{13}^\ast} \right)\left( {{s_{23}}{s_{12}} - {c_{23}}{c_{12}}\tilde s_{13}^\ast} \right) - \frac{B}{A}{t_{23}}{{\left( {{c_{23}}{s_{12}} + {s_{23}}{c_{12}}\tilde s_{13}^\ast} \right)}^2}} \right]{{\tilde m}_1} - \left( {{s_{23}}{c_{13}}{c_{23}}{c_{13}} + \frac{B}{A}{t_{23}}s_{23}^2c_{13}^2} \right){{\tilde m}_3}}}{{\left( {{c_{23}}{c_{12}} - {s_{23}}{s_{12}}\tilde s_{13}^\ast} \right)\left( { - {s_{23}}{c_{12}} - {c_{23}}{s_{12}}\tilde s_{13}^\ast} \right) + \frac{B}{A}{t_{23}}{{\left( {{c_{23}}{c_{12}} - {s_{23}}{s_{12}}\tilde s_{13}^\ast} \right)}^2}}},
\nonumber\\
\label{Eq:m2}
\end{eqnarray}
where ${\tilde s_{13}} = s_{13}e^{i\delta}$.  From this requirement, an important constraint on $B/A$, thereby, $\theta_{SC}$, is derived.  For the normal mass hierarchy, since $m^2_2 \ll m^2_3$, we must adjust $B/A$ to satisfy
\begin{eqnarray}
{{s_{23}}{c_{13}}{c_{23}}{c_{13}} + \frac{B}{A}{t_{23}}s_{23}^2c_{13}^2} \approx 0,
\label{Eq:m2_normal}
\end{eqnarray}
leading to
\begin{eqnarray}
\frac{{{{\cos }^2}{\theta _{SC}}{t_{23}}c_{13}^2}}{{{{\cos }^2}{\theta _{SC}} + {{\sin }^2}{\theta _{SC}}t_{23}^4}} \approx 0,
\label{Eq:m2_normal_SC}
\end{eqnarray}
from which we obtain that $\cos^2\theta_{SC}\approx 0$.  For the inverted mass hierarchy, since $m^2_1 \approx m^2_2$ ($\gg m^2_3$), we must adjust $B/A$ to satisfy
\begin{eqnarray}
&&
\left( {{c_{23}}{s_{12}} + {s_{23}}{c_{12}}\tilde s_{13}^\ast } \right)\left( {{s_{23}}{s_{12}} - {c_{23}}{c_{12}}\tilde s_{13}^\ast } \right) - \frac{B}{A}{t_{23}}{\left( {{c_{23}}{s_{12}} + {s_{23}}{c_{12}}\tilde s_{13}^\ast } \right)^2} \approx
\nonumber\\
&&
\qquad
e^{i\eta} \left[ {\left( {{c_{23}}{c_{12}} - {s_{23}}{s_{12}}\tilde s_{13}^\ast } \right)\left( { - {s_{23}}{c_{12}} - {c_{23}}{s_{12}}\tilde s_{13}^\ast } \right) + \frac{B}{A}{t_{23}}{{\left( {{c_{23}}{c_{12}} - {s_{23}}{s_{12}}\tilde s_{13}^\ast } \right)}^2}} \right],
\label{Eq:m2_inverted}
\end{eqnarray}
where $\eta$ is an arbitrary phase, which is reduced to
\begin{eqnarray}
&&
\left( {1 - \frac{B}{A}} \right)\left( {s_{12}^2 + e^{i\eta} c_{12}^2} \right){c_{23}}{s_{23}} + ({\rm terms~proportional~to}~s_{13})\approx 0,
\label{Eq:m2_inverted_result}
\end{eqnarray}
leading to $A\approx B$ that requires that $\sin^2\theta_{SC}\approx 0$.

Now, we are ready to discuss effects of CP-violation based on $M_{scaling}$.
\begin{enumerate}
\item To estimate Majorana phases $\varphi_{1,2,3}$ as well as $\phi_{2,3}$, we use Eq.(\ref{Eq:m2}) and choose $m_1=0$ for the normal mass hierarchy and $m_3=0$ for the inverted mass hierarchy to make arguments simpler. Majorana phases are calculable at $\theta_{13}=0$. 
\begin{enumerate}
\item For the normal mass hierarchy, when $m_1=0$, Eq.(\ref{Eq:m2}) becomes
\begin{eqnarray}
{\tilde m}_2 = - \frac{{\frac{{{{\cos }^2}{\theta _{SC}}{t_{23}}c_{13}^2}}{{{{\cos }^2}{\theta _{SC}} + {{\sin }^2}{\theta _{SC}}t_{23}^4}}}}{{\left( {{c_{23}}{c_{12}} - {s_{23}}{s_{12}}\tilde s_{13}^\ast } \right)\left( { - {s_{23}}{c_{12}} - {c_{23}}{s_{12}}\tilde s_{13}^\ast } \right) + \frac{B}{A}{t_{23}}{{\left( {{c_{23}}{c_{12}} - {s_{23}}{s_{12}}\tilde s_{13}^\ast } \right)}^2}}}{{\tilde m}_3},
\label{Eq:m2VSm3_normal}
\end{eqnarray}
whose denominator turns out to be $- c_{12}^2/t_{23}$+(terms proportional to $s_{13}$) for $\cos^2\theta_{SC}\approx 0$, which suggests that $\varphi_2 \approx \varphi_3$ leading to
\begin{eqnarray}
\phi_2 \approx \phi_3.
\label{Eq:m2VSm3_normal_Majorana}
\end{eqnarray}

If $\theta_{13}=0$, $\varphi _{2,3}$ can be calculated.  The requirement of $m_1=0$ yields ${\tilde m_2} = m/c_{12}^2c_{23}^2$ and $\tilde m_3 \approx M_{\mu \mu }/s_{23}^2$. From Eq.(\ref{Eq:MassRef}) with $\cos^2\theta_{SC}\approx 0$ for $m$, we reach
\begin{eqnarray}
{\varphi _2} =  - \arg \left( {{M_{\mu \mu}}} \right) \approx {\varphi _3}.
\label{Eq:normal_Majorana_phases}
\end{eqnarray}
\item For the inverted mass hierarchy, when $m_3=0$, Eq.(\ref{Eq:m2}) do not yield any positive results for  Majorana phases because they depend on $\delta$, whose contribution varies with $\sin^2\theta_{SC}$ contained in $B/A$.  For $\sin^2\theta_{SC}=0$, we obtain that
\begin{eqnarray}
{\tilde m}_2 = \frac{{1 + \frac{{{t_{23}}}}{{{t_{12}}}}\tilde s_{13}^\ast }}{{1 - {t_{23}}{t_{12}}\tilde s_{13}^\ast }}{{\tilde m}_1},
\label{Eq:m2VSm3_inverted}
\end{eqnarray}
which suggests that $\varphi_2 \approx \varphi_1$ leading to $\phi_2\approx 0$.
If $\theta_{13}=0$, $\varphi _{1,2}$ can be calculated. The requirement of $m_3=0$ yields $m = M_{\mu\mu}$ in Eq.(\ref{Eq:MassesIdeal}).  The simplest cases are ${\tilde m}_1 \approx {\tilde m}_2$ and ${\tilde m}_1 \approx -{\tilde m}_2$. For ${\tilde m}_1 \approx {\tilde m}_2$, 
\begin{eqnarray}
\phi_2\approx 0,
\label{Eq:m2VSm3_inverted_Majorana_theta_13=0}
\end{eqnarray}
should be realized at $\sin^2\theta_{SC}\approx 0$ as suggested from Eq.(\ref{Eq:m2VSm3_inverted}) and ${\tilde m}_1 \approx {M_{ee}}$ and ${\tilde m}_2 \approx M_{\mu \mu} /c_{23}^2$ are derived while, for ${\tilde m}_1 \approx -{\tilde m}_2$,
\begin{eqnarray}
\phi_2\approx \pm \pi,
\label{Eq:m2VSm3_inverted_Majorana_theta_13=0_2}
\end{eqnarray}
which should be realized at $\sin^2\theta_{SC} \neq 0$ and ${\tilde m}_1 \approx M_{ee}/\left(c_{12}^2 - s_{12}^2\right)$ and ${\tilde m}_2 \approx M_{\mu \mu }/c_{23}^2\left(c_{12}^2 - s_{12}^2\right)$  are derived. These results show that
\begin{eqnarray}
{\varphi _1} \approx  - \arg \left( {{M_{ee}}} \right),
\quad
{\varphi _2} \approx  - \arg \left( {{M_{\mu \mu }}} \right).
\label{Eq:inverted_Majorana_phases}
\end{eqnarray}
\end{enumerate}

\item CP-violating Dirac phase $\delta$ has been estimated to be \cite{DiracCPFormula}:
\begin{eqnarray}
\delta \approx \arg \left[ {\left( {\frac{1}{{{t_{23}}}}M_{\mu \tau }^\ast  + M_{\mu \mu }^\ast } \right)\delta {M_{e\tau }} + {M_{ee}}\delta M_{e\tau }^\ast  - {t_{23}}{M_{e\mu }}\delta M_{\tau \tau }^\ast } \right],
\label{Eq:DiracCP}
\end{eqnarray}
where an extra $\pi$ should be added for $c_{12}^2m_1^2 + s_{12}^2m_2^2 > m_3^2$, which is converted into
\begin{eqnarray}
\delta \approx \arg \left[ {\left( {1 - \frac{B}{A}} \right)M_{\mu \mu }^\ast \delta {M_{e\tau }} + {M_{ee}}\delta M_{e\tau }^\ast  - {t_{23}}{M_{e\mu }}\delta M_{\tau \tau }^\ast } \right].
\label{Eq:DiracCPConverted}
\end{eqnarray}
When conditions that realize the appropriate mass hierarchy are taken into account, Eq.(\ref{Eq:DiracCPConverted}) can be further simplified as follows:
\begin{enumerate}
\item For the normal mass hierarchy, since $\left|M_{\mu \mu }\right|\gg\left|M_{ee,e\mu}\right|$, Eq.(\ref{Eq:DiracCPConverted}) is reduced to
\begin{eqnarray}
\delta \approx \arg \left( {M_{\mu \mu }^\ast \delta {M_{e\tau }}} \right).
\label{Eq:DiracCPNormal}
\end{eqnarray}
\item For the inverted mass hierarchy, since $A\approx B$, Eq.(\ref{Eq:DiracCPConverted}) is reduced to
\begin{eqnarray}
\delta \approx \arg \left( {{M_{ee}}\delta M_{e\tau }^\ast  - {t_{23}}{M_{e\mu }}\delta M_{\tau \tau }^\ast } \right)+\pi.
\label{Eq:DiracCPInverted}
\end{eqnarray}
\end{enumerate}
It will be demonstrated that the simplified forms of $\delta$, Eqs.(\ref{Eq:DiracCPNormal}) and (\ref{Eq:DiracCPInverted}), numerically well reproduce actual values of $\delta$.
\end{enumerate}
These predictions are checked against numerical analysis.

\begin{figure}[t]
\begin{center}
\includegraphics*[10mm,230mm][300mm,295mm]{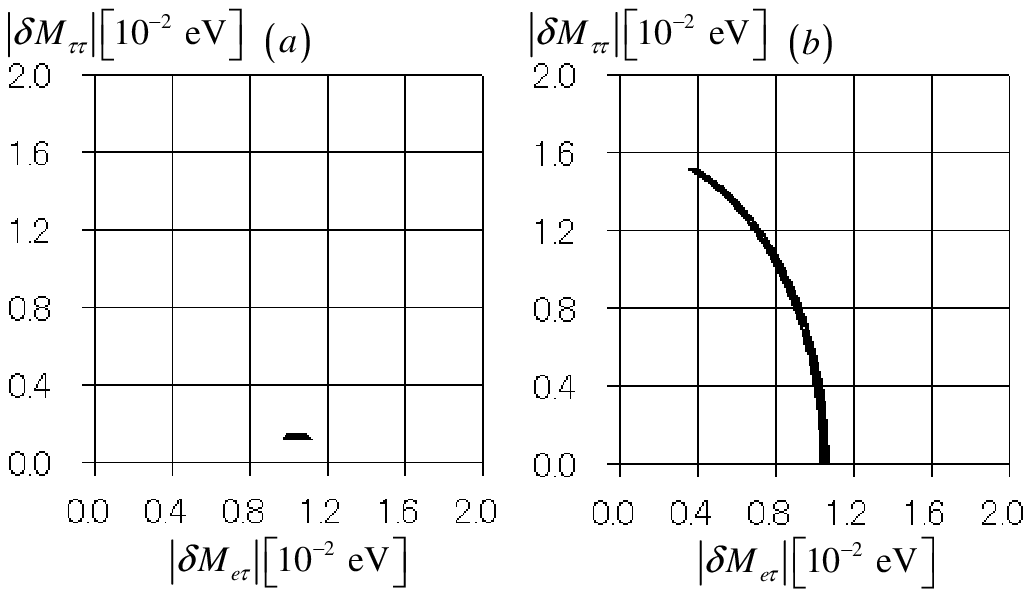}
\caption{The prediction of $\left|\delta M_{\tau\tau}\right|$ as a function of $\left|\delta M_{e\tau}\right|$ for (a) the normal mass hierarchy or (b) the inverted mass hierarchy.}
\label{Fig:vsDMetauDMtautau}
\includegraphics*[10mm,232mm][300mm,290mm]{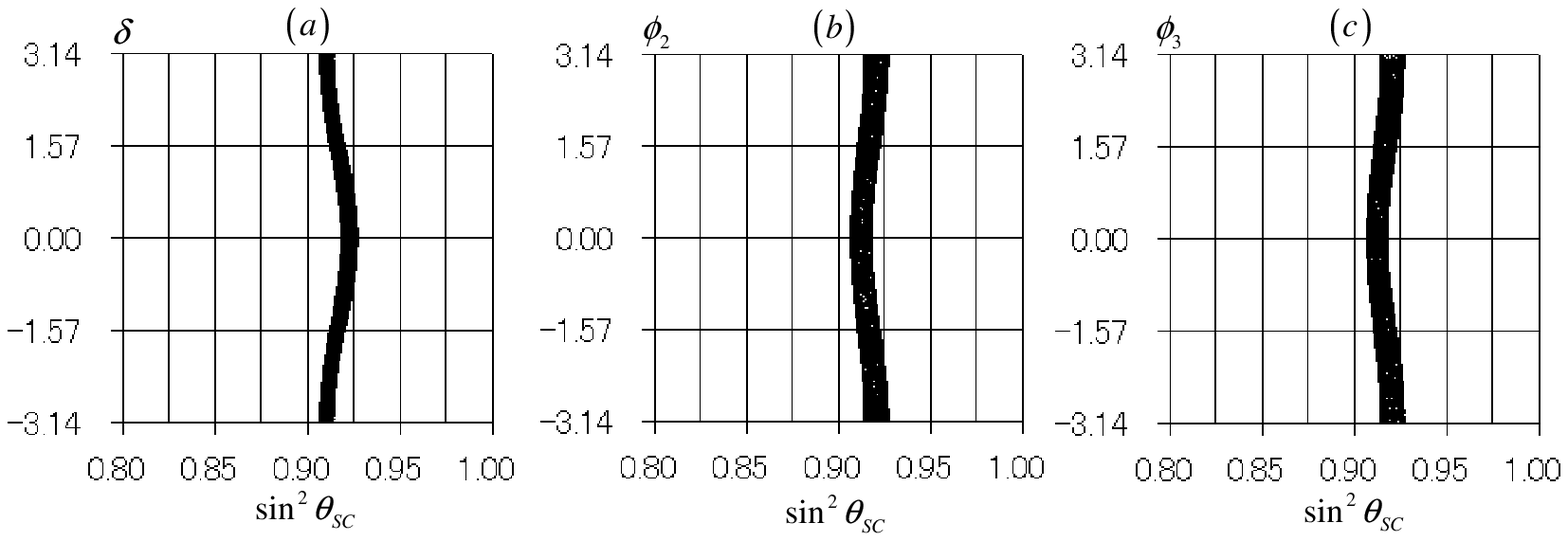}
\caption{The predictions of (a) $\delta$ or (b) $\phi_2$ and (c) $\phi_3$ as functions of $\sin^2\theta_{SC}$  for the normal mass hierarchy.}
\label{Fig:vsSinSC_Normal}
\end{center}
\end{figure}
\begin{figure}[t]
\begin{center}
\includegraphics*[10mm,232mm][300mm,295mm]{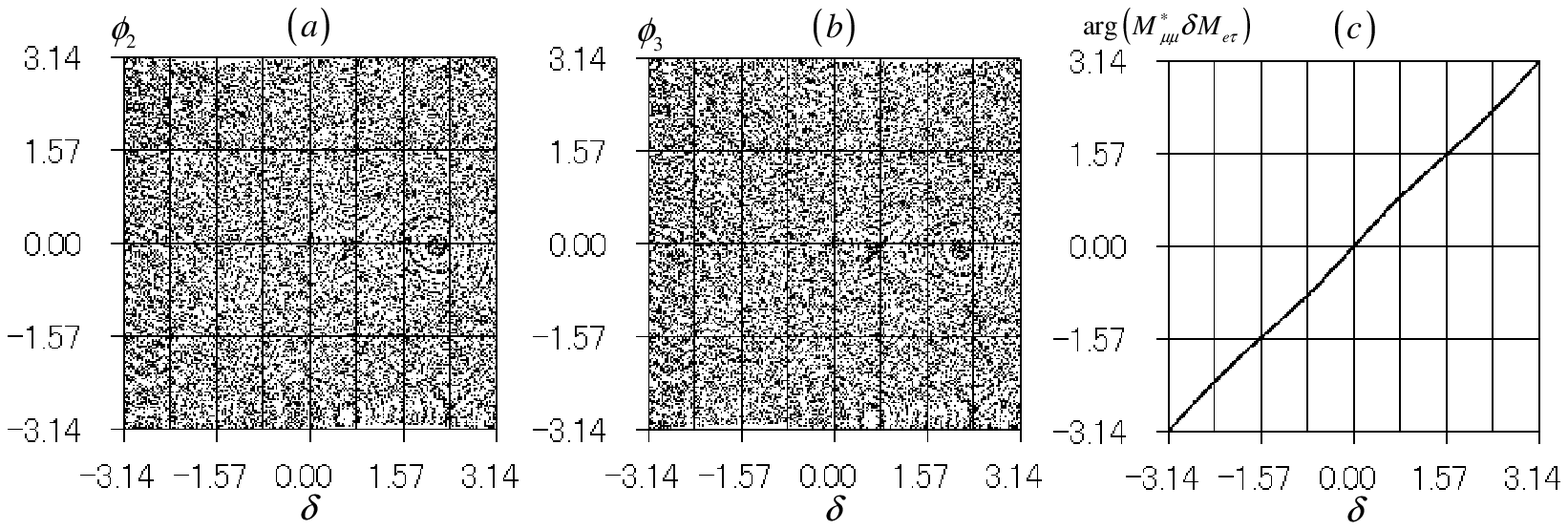}
\caption{The same as in FIG.\ref{Fig:vsSinSC_Normal} but for (a) $\phi_2$, (b) $\phi_3$ and (c) the simpler form of $\delta$ as functions of $\delta$.}
\label{Fig:vsDelta_Normal}
\includegraphics*[10mm,232mm][300mm,290mm]{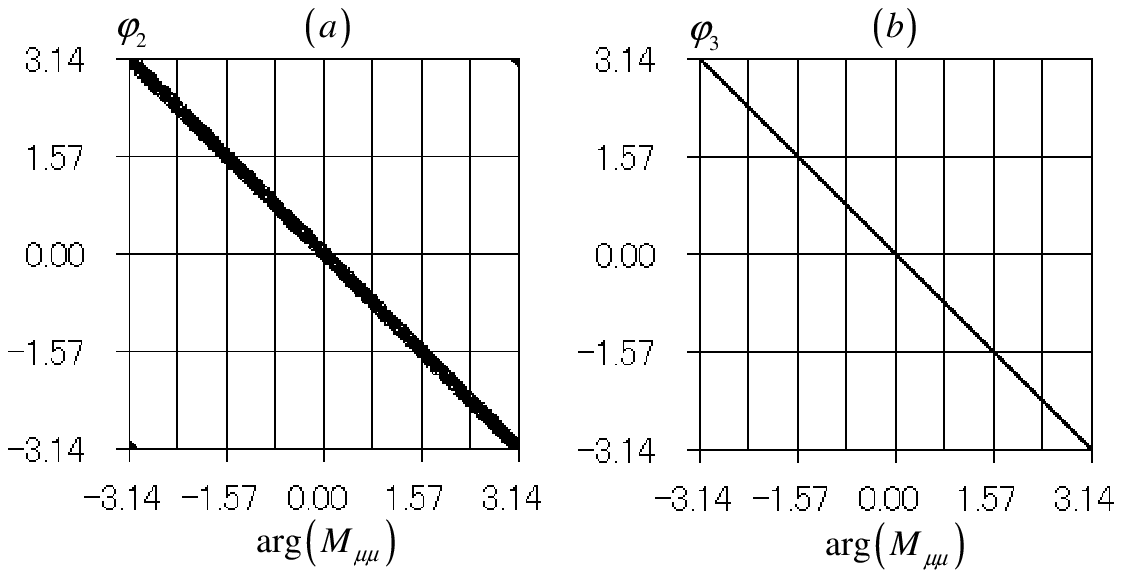}
\caption{The same as in FIG.\ref{Fig:vsSinSC_Normal} but for (a) $\varphi_2$ and (b) $\varphi_3$ as functions of $M_{\mu\mu}$.}
\label{Fig:vsMassPhases_Normal}
\end{center}
\end{figure}
\begin{figure}[t]
\begin{center}
\includegraphics*[10mm,232mm][300mm,295mm]{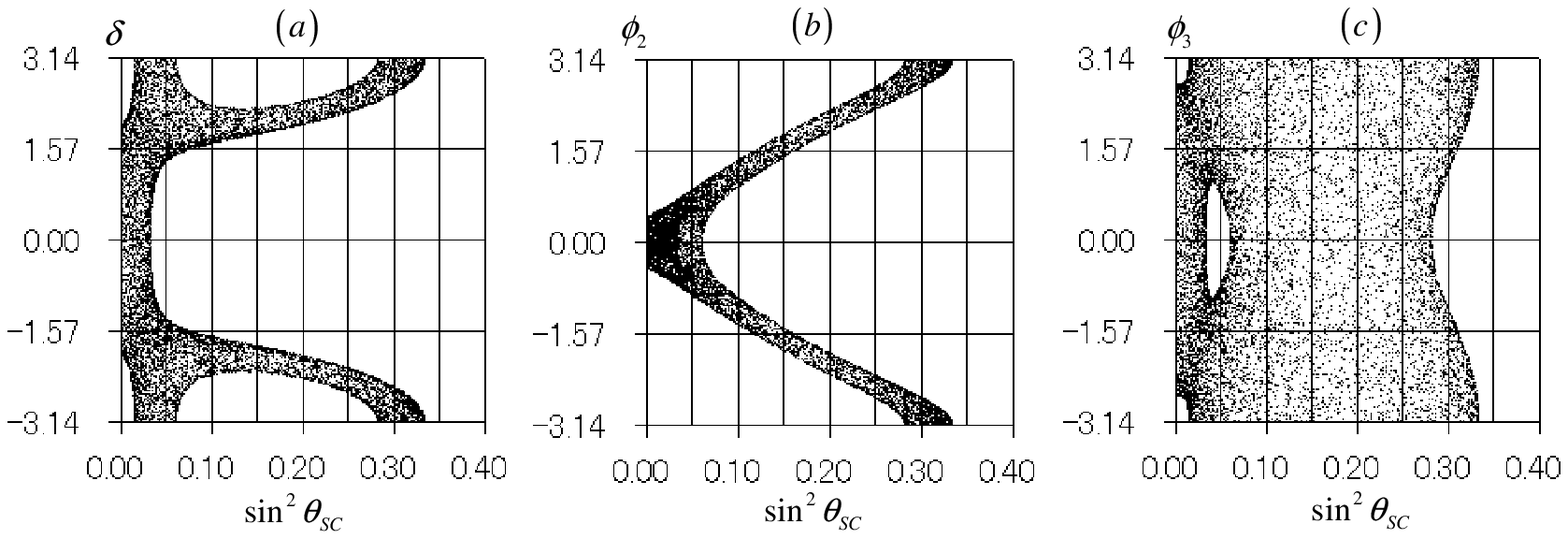}
\caption{The same as in FIG.\ref{Fig:vsSinSC_Normal} but for the inverted mass hierarchy.}
\label{Fig:vsSinSC_Inverted}
\includegraphics*[10mm,228mm][300mm,290mm]{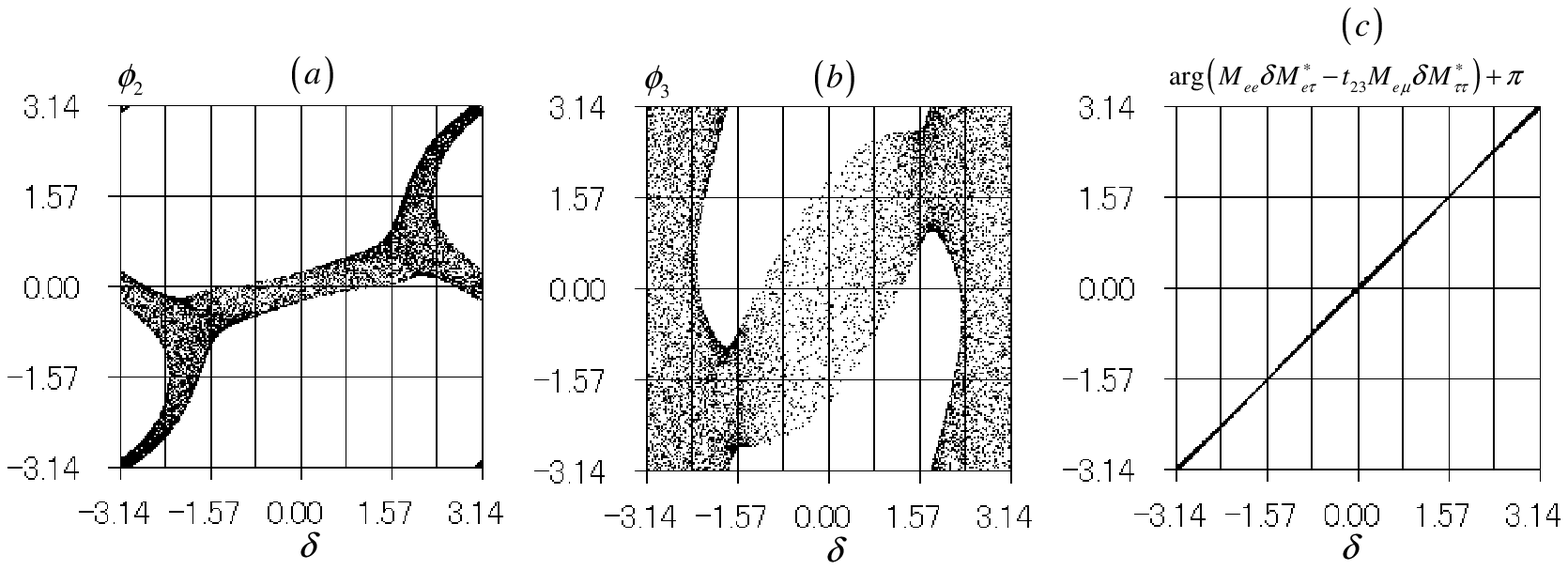}
\caption{The same as in FIG.\ref{Fig:vsDelta_Normal} but for the inverted mass hierarchy.}
\label{Fig:vsDelta_Inverted}
\end{center}
\end{figure}
\begin{figure}[t]
\begin{center}
\includegraphics*[10mm,232mm][300mm,290mm]{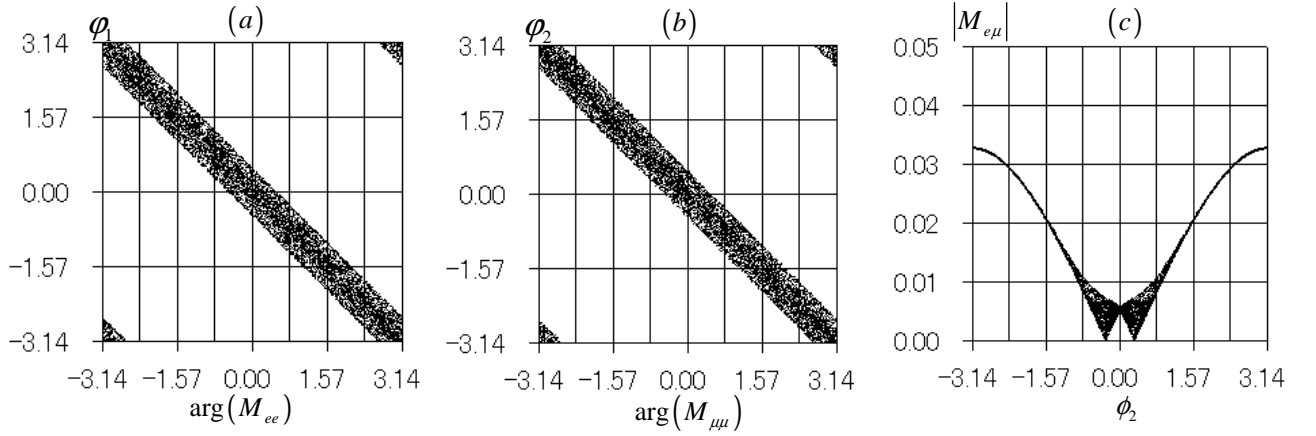}
\caption{The predictions of (a) $\varphi_1$ as a function of $M_{ee}$, (b) $\varphi_2$ as a function of $M_{\mu\mu}$ and (b) $\left|M_{e\mu}\right|$ as a function of $\phi_2(=\varphi_2-\varphi_1)$ for the inverted mass hierarchy.}
\label{Fig:vsMassPhases_Inverted}
\end{center}
\end{figure}
\begin{figure}[t]
\begin{center}
\includegraphics*[10mm,232mm][300mm,292mm]{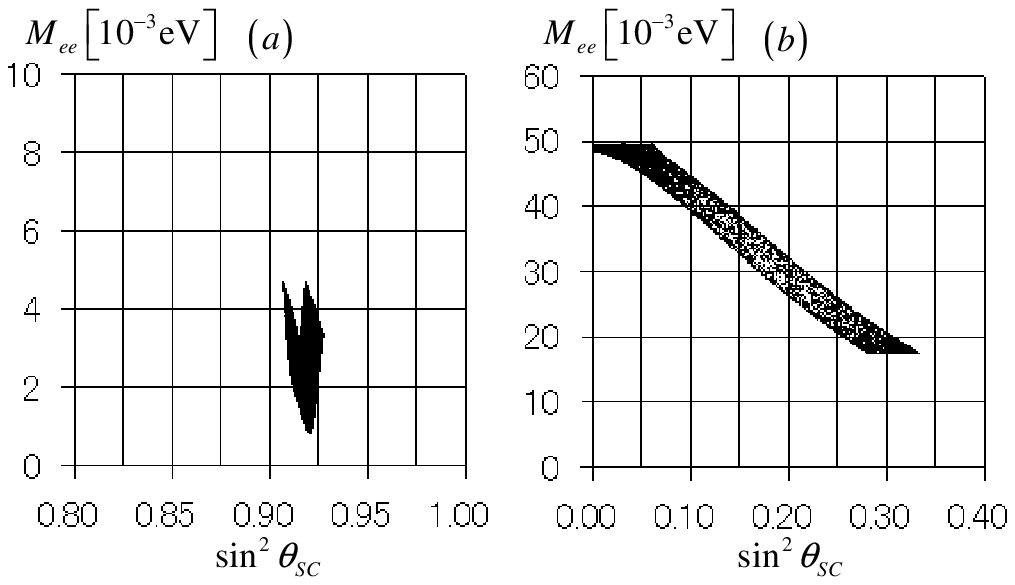}
\caption{The prediction of $\left|M_{ee}\right|$ as a function of $\sin\theta^2_{SC}$ for (a) the normal mass hierarchy or (b) the inverted mass hierarchy.}
\label{Fig:vsSinSCMee}
\end{center}
\end{figure}

\section{\label{sec:calculation} Numerical Calculations}
We have performed numerical calculations to estimate sizes of CP-violating phases and to discuss their possible correlations with other quantities. The parameters used are
\begin{eqnarray}
\Delta m^2_{21} ~[10^{-5}~{\rm eV}^2] & = &7.62,
\qquad
\Delta m^2_{31} ~[10^{-3}~{\rm eV}^2] = 2.55,
\label{NumericalMassSquared}\\
\sin ^2 \theta _{12} & = &0.32,
\qquad
\sin ^2 \theta _{23} =0.43,
\qquad
\sin ^2 \theta _{13} =0.025.
\label{NumericalAngle}
\end{eqnarray}
For the sake of simplicity, we adopt $m_1$=0.001 eV ($m_3$=0.001 eV) for the normal (inverted) mass hierarchy.   Our results are listed in FIG.\ref{Fig:vsDMetauDMtautau}-FIG.\ref{Fig:vsMassPhases_Inverted}. The size of the breaking terms $\delta M_{e\tau,\tau\tau}$ is plotted in FIG.\ref{Fig:vsDMetauDMtautau}.  The dependence of CP-violating phases on $\sin^2\theta_{SC}$ is shown in FIG.\ref{Fig:vsSinSC_Normal} and FIG.\ref{Fig:vsSinSC_Inverted}.  Shown in FIG.\ref{Fig:vsDelta_Normal} and FIG.\ref{Fig:vsDelta_Inverted} are predictions on the dependence of (a) the simpler forms of $\delta$, (b) $\phi_2$ and (c) $\phi_3$ on $\delta$.  The remaining figures, FIG.\ref{Fig:vsMassPhases_Normal} and FIG.\ref{Fig:vsMassPhases_Inverted}, are predictions on the dependence of $\arg \left( M_{ee,e\mu,\mu \mu }\right)$ on Majorana phases.

Form these figures, we can observe the following property of CP-violating phases:
\begin{itemize}
\item For the normal mass hierarchy, where $\theta_{SC}$ is restricted to the small range of $0.91 \lesssim\sin^2\theta_{SC}\lesssim 0.93$ (FIG.\ref{Fig:vsSinSC_Normal}) and $\sin^2\theta_{SC}=1$ is forbidden owing to the constraint of $m^2_2>m^2_1$,
\begin{itemize} 
\item $\left|\delta M_{\tau\tau}\right|$ is well suppressed to confirm the validity of Eq.(\ref{Eq:DiracCPNormal}) and $\left|\delta M_{e\tau}\right|\sim 0.01$ eV (FIG.\ref{Fig:vsDMetauDMtautau} (a));
\item $\left|\delta\right|\rightarrow \pi$ as $\sin^2\theta_{SC}$ reaches smaller values (FIG.\ref{Fig:vsSinSC_Normal} (a));
\item $\phi_2 \approx\phi_3$ (FIG.\ref{Fig:vsSinSC_Normal} (b) and (c));

\item $\phi_2$ and $\phi_3$ exhibit no correlation with $\delta$ (FIG.\ref{Fig:vsDelta_Normal} (a) and (b));
\item $\arg \left( {M_{\mu \mu }^\ast \delta {M_{e\tau }}} \right)$ is almost identical to $\delta$ (FIG.\ref{Fig:vsDelta_Normal} (c));
\item $\varphi_2 \approx\varphi_3\approx -\arg\left(M_{\mu\mu}\right)$ (FIG.\ref{Fig:vsMassPhases_Normal} (a) and (b)).
\end{itemize}
\item For the inverted mass hierarchy, where $\sin^2\theta_{SC}\lesssim 0.33$ (FIG.\ref{Fig:vsSinSC_Inverted}),
\begin{itemize} 
\item $\left|\delta M_{e\tau}\right|$ and $\left|\delta M_{\tau\tau}\right|$ are comparable and combined contributions amount to 0.01-0.02 eV (FIG.\ref{Fig:vsDMetauDMtautau} (b));
\item  $\left|\delta\right|$ ($\left|\phi_2\right|$) is larger than $\pi/2$ for $\sin^2\theta_{SC}\gtrsim 0.1$ ($\sin^2\theta_{SC}\gtrsim 0.15$) and tends to be larger as $\sin^2\theta_{SC}$ gets larger  (FIG.\ref{Fig:vsSinSC_Inverted} (a),(b));
\begin{itemize} 
\item $\left|\delta\right|\gtrsim \pi/2$ for $\sin^2\theta_{SC}\gtrsim 0.1$ (FIG.\ref{Fig:vsSinSC_Inverted} (a));
\item $\left|\phi_2\right|\rightarrow \pi$ as $\sin^2\theta_{SC}\rightarrow 0.33$, where $\left|\phi_2\right|\lesssim 0.5$ for $\sin^2\theta_{SC}\approx 0$ and $\left|\phi_2\right|\approx \pi$ for $\sin^2\theta_{SC}\approx 0.33$ (FIG.\ref{Fig:vsSinSC_Inverted} (b));
\end{itemize} 
\item $\phi_3$ does not exhibit no clear correlation with $\sin^2\theta_{SC}$ (FIG.\ref{Fig:vsSinSC_Inverted} (c));
\item $\phi_2$ is scattered around the straight line of $\phi_2=\delta/4$ for $\left|\delta\right|\lesssim \pi/2$ while $\phi_2=0,\pi$ for $\left|\delta\right|\approx \pi$ (FIG.\ref{Fig:vsDelta_Inverted} (a));
\item $\arg \left( {{M_{ee}}\delta M_{e\tau }^\ast  - {t_{23}}{M_{e\mu }}\delta M_{\tau \tau }^\ast } \right)+\pi$ is almost identical to $\delta$ (FIG.\ref{Fig:vsDelta_Inverted} (c));
\item $\varphi_1$ is scattered around the line $\varphi_1 = -\arg\left(M_{ee}\right)$ (FIG.\ref{Fig:vsMassPhases_Inverted} (a)) while $\varphi_2$ is scattered around the line $\varphi_2 = -\arg\left(M_{\mu\mu}\right)$ (FIG.\ref{Fig:vsMassPhases_Inverted} (b));
\item $\left|\phi_2\right|\propto \left|M_{e\mu}\right|$, where $\phi_2$ is located around 0 leading to $m_1e^{-i\varphi_1}\approx m_2e^{-i\varphi_2}$ if $\left|M_{e\mu}\right|$ is suppressed while $\phi_2$ is located around $\pm\pi$  leading to $m_1e^{-i\varphi_1}\approx -m_2e^{-i\varphi_2}$ if $\left|M_{e\mu}\right|$ is enhanced (FIG.\ref{Fig:vsMassPhases_Inverted} (c));
\end{itemize}
\end{itemize}
Some of these results of Majorana phases can be explained by the predictions made at $\sin\theta_{13}=0$:
\begin{itemize} 
\item For the normal mass hierarchy with $m_1=0$, 
\begin{itemize} 
\item $\phi_2 \approx \phi_3$ from Eq.(\ref{Eq:m2VSm3_normal_Majorana});
\item $\phi_2 \approx \phi_3$ based on ${\varphi _2} =  - \arg \left( {{M_{\mu \mu}}} \right) \approx {\varphi _3}$ from Eq.(\ref{Eq:normal_Majorana_phases}).
\end{itemize} 
\item For the inverted mass hierarchy with $m_3=0$, 
\begin{itemize} 
\item ${\varphi _1} \approx  - \arg \left( {{M_{ee}}} \right)$ and ${\varphi _2} \approx  - \arg \left( {{M_{\mu \mu }}} \right)$ from Eq.(\ref{Eq:inverted_Majorana_phases});
\end{itemize} 
\end{itemize}
The result of FIG.\ref{Fig:vsMassPhases_Inverted} (c) about the behavior of $\phi_2$ can be understood in the following way: Eqs.(\ref{Eq:m2VSm3_inverted_Majorana_theta_13=0}) and (\ref{Eq:m2VSm3_inverted_Majorana_theta_13=0_2}) to predict sizes of $\phi_2$ are rephrased in terms of $M_{e\mu}$. Since ${\tilde m}_{1,2}$ in Eq.(\ref{Eq:MassesIdeal}) are transformed into
\begin{eqnarray}
{\tilde m}_1 = \frac{1}{2}\left( {\frac{{{M_{\mu \mu }}}}{{c_{23}^2}} + {M_{ee}} - \frac{{2{M_{e\mu }}}}{{{c_{23}}\sin 2{\theta _{12}}}}} \right),
\quad
{\tilde m}_2 = \frac{1}{2}\left( {\frac{{{M_{\mu \mu }}}}{{c_{23}^2}} + {M_{ee}} + \frac{{2{M_{e\mu }}}}{{{c_{23}}\sin 2{\theta _{12}}}}} \right),
\label{Eq:MassesIdealAlt}
\end{eqnarray}
we obtain that ${\tilde m_1}\approx{\tilde m_2}$ leading to $\phi_2\approx 0$ when $\left|M_{e\mu}\right|$ is suppressed and that  ${\tilde m_1}\approx-{\tilde m}_2$ leading to $\phi_2\approx\pm\pi$ when $\left|M_{e\mu}\right|$ is enhanced.  Then, we observe that
\begin{itemize} 
\item $\phi_2\approx 0$ for the suppressed $\left|M_{e\mu}\right|$;
\item $\phi_2\approx \pm \pi$ for the enhanced $\left|M_{e\mu}\right|$.
\end{itemize} 
and both cases are smoothly connected as indicated by FIG.\ref{Fig:vsMassPhases_Inverted} (c).

Finally, we plot $\left|M_{ee}\right|$ to be measured by neutrinoless double beta decay \cite{DoubleBeta} as a function of $\sin^2\theta_{SC}$ in FIG.\ref{Fig:vsSinSCMee}.  In the normal mass hierarchy, $\left|M_{ee}\right|$ should be suppressed to yield the suppressed $m_1$ and is estimated to be 0.001 eV-0.004 eV.  On the other hand, in the inverted mass hierarchy, since $\left|M_{ee}\right|$ reflects the size of $m_{1,2}$ of order ${\mathcal O}(\sqrt{\Delta m^2_{atm}})$, we expect that $\left|M_{ee}\right|\approx$0.05 eV.  In FIG.\ref{Fig:vsSinSCMee} (b), $\left|M_{ee}\right|$ is estimated to be $0.02-0.05$ eV.  It seems that $\left|M_{ee}\right|$ is scattered around the line of $\left|M_{ee}\right|=0.055-0.11\sin^2\theta_{SC}$ [eV].

\section{\label{sec:summary} Summary and Discussions}
In summary, we have proposed the generalized scaling anzatz for flavor neutrino masses, which dictates that
\begin{eqnarray}
\frac{{{M_{i\tau }}}}{{{M_{i\mu }}}} =  - {\kappa _i}{t_{23}}~\left(i=e,\mu,\tau\right)
\label{Eq:ScalingRule-summary}
\end{eqnarray}
with $\left(\kappa_e, \kappa_\mu, \kappa_\tau\right)$=$\left( 1,B/A,1/B\right)$, where $A = {\cos ^2}{\theta _{SC}} + {\sin ^2}{\theta _{SC}}t_{23}^4$ and $B = {\cos ^2}{\theta _{SC}} - {\sin ^2}{\theta _{SC}}t_{23}^2$, which gives $\theta_{13}=0$.  The angle $\theta_{SC}$ is defined by ${\sin }^2{\theta _{SC}} = {{c_{23}^2\left( {{M_{\mu \tau }} + {t_{23}}{M_{\mu \mu }}} \right)}}/[\left( {1 - t_{23}^2} \right){M_{\mu \tau }} + {t_{23}}M_{\mu \mu }]$ and can be rephrased in terms of seesaw textures parameterized by $a_1={b_3} + {t_{23}}{b_2} = {c_3} + {t_{23}}{c_2} = 0$ with (a) $a_3 = -t_{23}a_2$ as in Eq.(\ref{Eq:Solution-MSeesaw1}) leading to $\sin^2\theta_{SC}=0$ and with (b) $a_3 = a_2/t_{23}$ as in Eq.(\ref{Eq:Solution-MSeesaw2}) leading to $\sin^2\theta_{SC}= - a_3^2/(M_1 {{M_{\tau \tau }}})$.  To induce $\theta_{13}\neq 0$, the breaking terms $\delta M_{e\tau}$ for $M_{e\tau}$ and $\delta M_{\tau\tau}$ for $M_{\tau\tau}$ should be included. When these breaking terms are minimal ones that necessarily induce $\theta_{13}\neq 0$ and Dirac CP-violation, the constraint of ${M_{\mu \tau }} + Bt_{23}M_{\mu \mu }/A=0$ is satisfied.  At the same time, this constraint ensures that the definition of $\theta_{SC}$ is valid even at $\theta_{13}\neq 0$.  We have found that 
\begin{itemize}
\item CP-violating Dirac phase can be estimated to be ${\delta  \approx \arg \left( {M_{\mu \mu }^\ast\delta {M_{e\tau }}} \right)}$ for the normal mass hierarchy and to be ${\delta  \approx \arg \left( {{M_{ee}}\delta M_{e\tau }^\ast - {t_{23}}{M_{e\mu }}\delta M_{\tau \tau }^\ast} \right) + \pi }$ for the inverted mass hierarchy;
\item CP-violating Majorana phases can be estimated from ${\varphi _2} \approx {\varphi _3} \approx  - \arg \left( {{M_{\mu \mu }}} \right)$ for the normal mass hierarchy and from ${\varphi _1} \approx  - \arg \left( {{M_{ee}}} \right)$ and ${\varphi _2} \approx  - \arg \left( {{M_{\mu \mu }}} \right)$ for the inverted mass hierarchy. 
\end{itemize}

From the numerical calculation, where the lightest neutrino mass is taken to be 0.001 eV, we have observed that the approximate scaling anzatz describes the normal mass hierarchy for $0.91 \lesssim\sin^2\theta_{SC}\lesssim 0.93$ and the inverted mass hierarchy for $\sin^2\theta_{SC}\lesssim 0.33$ and that breaking terms of the general scaling anzatz are estimated to be at most around ${\mathcal O}$(0.01) eV. The effective mass to be measured in neutrinoless double beta decay $\left|M_{ee}\right|$ is found to be 0.001-0.004 eV (0.02 eV-0.05 eV) in the normal (inverted) mass hierarchy. The following numerical properties are also found:
\begin{itemize}
\item For the normal mass hierarchy, 
\begin{itemize}
\item $\left|\delta\right|$ tends to be smaller as $\sin^2\theta_{SC}$ gets larger while $\left|\phi_{2,3}\right|$ tends to be larger as $\sin^2\theta_{SC}$ gets larger;
\item there is no correlation between $\delta$ and $\phi_{2,3}$.
\end{itemize}
\item For the inverted mass hierarchy,
\begin{itemize}
\item  $\left|\delta\right|$ is larger than $\pi/2$ for $\sin^2\theta_{SC}\gtrsim 0.1$  while $\left|\phi_2\right|$ is larger than $\pi/2$ for $\sin^2\theta_{SC}\gtrsim 0.15$ and both tend to be larger as $\sin^2\theta_{SC}$ gets larger;
\item there is a clear correlation between $\delta$ and $\phi_2$ that shows $\phi_2$ is scattered around the straight line of $\phi_2=\delta/4$ for $\left|\delta\right|\lesssim \pi/2$ while $\phi_2=0,\pi$ for $\left|\delta\right|\approx \pi$;
\item the proportionality of $\left|\phi_2\right|\propto \left|M_{e\mu}\right|$  symbolizes the behavior of $\left|\phi_2\right|$;
\item $\left|M_{ee}\right|$ is scattered around the line of $\left|M_{ee}\right|=0.055-0.11\sin^2\theta_{SC}$ [eV].
\end{itemize}
\end{itemize}
Since the generalized scaling anzatz is compatible with the seesaw mechanism, we may discuss the creation of the baryon asymmetry of the Universe via the leptogenesis \cite{Leptogenesis} considering results of the generalized scaling anzatz.  This subject will be discussed elsewhere \cite{NewResearch}.
 
\vspace{3mm}
\noindent
\centerline{\small \bf ACKNOWLEGMENTS}

The author would like to thank T. Kitabayashi for reading manuscript and useful comments.



\end{document}